\documentclass[letter,scriptaddress,twocolumn]{revtex4}

	\usepackage{amsmath}
	\usepackage[sort&compress]{natbib}
	\usepackage{makeidx}
	\usepackage{amsfonts}
	\usepackage[ansinew]{inputenc}
	\usepackage[usenames,dvipsnames]{pstricks}
	\usepackage{subfigure}
	\usepackage{epsfig}
	\usepackage{pst-grad} 
	\usepackage{pst-plot} 
	\usepackage[colorlinks,hyperindex]{hyperref}
	\hypersetup
	{
		colorlinks,%
		citecolor=blue,%
		linkcolor=blue,%
		urlcolor=black,%
	}

\newcommand{\beqn}{\begin{equation}}
\newcommand{\eeqn}{\end{equation}}
\newcommand{\beqna}{\begin{eqnarray}}
\newcommand{\eeqna}{\end{eqnarray}}
	
\makeindex

\begin{document}
\title{Azimuthal anisotropy in particle distribution in A Multi-phase Transport Model}
\author{Soumya Sarkar$^{1,2}$, Provash Mali$^{1}$ and Amitabha Mukhopadhyay$^{1}$}
\email{amitabha$_$62@rediffmail.com}
\affiliation{$^{1}$Department of Physics, University of North Bengal, Siliguri 734013, West Bengal, India\\
$^{2}$Department of Physics, Siliguri College, Siliguri 734001, West Bengal, India}
\begin{abstract}
Anisotropic flow of hadronic matter is considered as a sensitive tool to detect the early stage dynamics of high-energy heavy-ion collisions. Taking the event by event fluctuations of the collision geometry into account, the elliptic flow parameter and the triangular flow parameter derived from the azimuthal distribution of produced hadrons, are investigated within the framework of a multiphase transport (AMPT) model, at a collision energy that in near future will typically be available at the Facility for Antiproton and Ion Research. The dependence of elliptic and triangular flow parameters on initial fluctuations, on parton scattering cross-sections, their mass ordering on different hadron species and on the constituent quark number scaling are examined. The AMPT simulation can not exactly match the elliptic flow results on Pb + Pb collision at 40A GeV of the NA49 experiment. The simulation results presented in this work are expected to provide us with an insight to study flow properties at high baryonic density but moderate temperature, and also with an opportunity to compare similar results available from RHIC and LHC experiments.\\\\
PACS number(s): 25.75.Nq, 24.10.Lx, 25.75.Ld, 25.75.Gz
\end{abstract}
\maketitle
\section{Introduction}
\label{intro}
The study of azimuthal anisotropy of final state hadrons is believed to be one of the important tools that can extract significant information regarding particle interactions in a hot and dense nuclear and/or partonic medium produced in high-energy heavy-ion collisions. Properties of this kind of matter are widely believed to be guided by the rules of quantum chromodynamics (QCD). Several important results on the collective behavior of final state particles have already been obtained by using the Fourier decomposition of their azimuthal distributions. Amongst all the second harmonic coefficient, also known as the elliptic flow parameter $(v_2)$, is of special interest \cite{Voloshin08}. Large $v_2$ values observed in Relativistic Heavy-ion Collider (RHIC) \cite{Ackermann01, Adler01, Adler03, Adare07} and Large Hadron Collider (LHC) \cite{Aamodt10, Aamodt11, Aad12, Chatrchyan12} experiments are understood to be due to a strongly interacting nature of the extended QCD state composed of loosely coupled quarks and gluons. The $v_2$ parameter is sensitive to the equation of state, transport properties of the medium, degree of thermalization achieved by the system, and also to the initial conditions of a collision \cite{Sorge97, Ollitrault92, Romatschke07, Luzum08, Huovinen01}. At low transverse momentum $(p_{_T})$ a mass ordering of $v_2$ with respect to different hadron species, and a scaling with respect to the number of constituent quarks (NCQ) that the hadron under consideration is made of, have been observed both in RHIC \cite{Adams05, Abelev07, Abelev10} and in LHC experiments \cite{Abelev15}. The NCQ scaling enables us to understand how significant the partonic degrees of freedom are, in the intermediate `fireball' created in any high-energy nucleus-nucleus $(AB)$ collision \cite{Molnar03,Dong04}.\\\\ 
In recent years the third harmonic coefficient $v_3$ of the Fourier decomposition of the azimuthal distribution, also called the triangular flow parameter, has gained attention and has also been studied extensively \cite{Alver10a,Teaney11,Gavin12}. Originally it was perceived though, that due to a left-right symmetry prevailing in the transverse plane of a collision, the contribution from odd harmonics to the particle azimuthal distribution would vanish. However, now it is widely accepted that the event-by-event fluctuating position of the nucleons participating in an $AB$ collision, often assumes a triangular shape (preferably called the triangularity), which with the evolution of the interacting system is converted into a momentum space anisotropy. Triangular flow is sensitive to the correlations present in the early stage of the $AB$ collision, and it has been proposed that the triangular anisotropy can explain the near side `ridge' and the away side `shoulder' structures present in two-particle (dihadron) azimuthal correlations \cite {Alver10a}. Furthermore, triangular flow is also believed to be sensitive to the viscous effects of the `fireball' medium as suggested by some simulation studies on relativistic viscous hydrodynamics \cite{Alver10b, Schenke}. At high temperature and low baryon density, triangular flow of produced hadrons has been studied as a function of $p_{_T}$, pseudorapidity $(\eta)$, centrality (often measured in terms of the number of participating nucleons $N_{\rm part}$), and triangularity $(\varepsilon_3)$ \cite{Adamczyk13, Aamodt13, Solanki12, Sun14, Sun15, Han11}. But the effect of the aforementioned initial fluctuations on final state azimuthal anisotropy is not yet fully explored at low and moderate collision energies. The upcoming Compressed Baryonic Matter (CBM) experiment \cite{CBM} to be undertaken at the Facility for Anti-proton and Ion Research (FAIR), is dedicated to study the color deconfined QCD matter at low to moderate temperature and high baryon density. The CBM will be a fixed target experiment on $AB$ interactions where the proposed incident beam energy will be in the range $E_{\rm lab} \sim 10-40$ GeV per nucleon. At such interaction energies it is expected that a baryon density $\rho_{_B}\sim 6-12$ times the normal nuclear matter density will be created in the central rapidity region \cite{Stocker86}. As far as high-energy $AB$ interactions are concerned, the CBM experiment will be complementary to the ongoing RHIC and LHC programs.\\\\
As mentioned above, the harmonic flow co-efficients $(v_n)$ of different order $(n)$, or each type of anisotropy can be obtained from the Fourier expansion of the azimuthal distribution of produced particles, the azimuthal angle being measured with respect to that of a participant plane angle $(\psi_n)$ \cite{Poskanzer98}. The azimuthal angle distribution of particles can be Fourier decomposed as \cite{Han11}, 
\beqn
\frac{dN}{d\phi} \propto \left[1 + 2\sum_{n=1}^{\infty} v_n\, \cos\left\{n(\phi - \psi_n)\right\}\right]
\eeqn
where $\phi$ is the momentum azimuthal angle of each particle, and $\psi_n$ is the azimuthal angle of the participant plane associated with the $n$-th harmonic that maximizes the eccentricity of the participating nucleons. In the center of mass system of the participating nucleons $\psi_n$ is given by,
\beqn
\psi_n = \frac{1}{n}\left[\arctan \frac{\left<r^2\,\sin(n\varphi)\right>}{\left<r^2\,\cos(n\varphi)\right>} + \pi\right]
\eeqn
where $(r,\,\varphi)$ denote the position co-ordinates of participating nucleons in a plane polar system and $\left<~\right>$ denotes a density weighted average over the initial states. As the number and position co-ordinates of participating nucleons fluctuate from one Au + Au collision to the other, that is going to affect the $\psi_n$, and therefore, the $v_n$ values. On the other hand the initial geometric deformation of the overlapping region of two colliding nuclei is quantified by, 
\beqn
\varepsilon_n = \frac{\sqrt{\left<r^2\cos(n\varphi)\right>^2 + \left<r^2\sin(n\varphi)\right>^2} }{\left <r^2 \right>}
\label{en}
\eeqn
Taking the effects of initial fluctuations into account the anisotropic flow parameter $v_n$ is defined as, 
\beqn
v_n = \left <\cos[n(\phi - \psi_n)] \right>
\label{vn}
\eeqn
Our present understanding of the dynamics of partonic and/or hadronic matter produced in $AB$ collisions around FAIR energy lacks experimental evidence. Under such circumstances we have to rely upon model calculations and Monte Carlo Simulations built thereof. For all practical purposes, simulation codes that can describe the nature of global variables associated with multiparticle production in high-energy interactions with reasonable success, should be chosen for a more in depth study. Very recently we have reported some such simulated results on the centrality dependence of elliptic flow parameter, and some aspects of kinetic radial flow of charged hadrons at FAIR energies \cite{Sarkar}. However, due importance to issues like initial fluctuations and parton degrees of freedom are not given in that analysis. The main objective of this article is therefore, to use a set of simulated multiparticle emission data on Au + Au collision at $E_{\rm lab}=30A$ GeV, and study the azimuthal anisotropy of charged hadrons in the final state, where the effects of event-to-event initial fluctuations and binary (partonic) scattering are taken into account. From statistical considerations \cite{Randrup06} it has been shown that in fixed target $AB$ experiments, at and around $E_{\rm lab}=30A$ GeV the expected net baryon density in the intermediate `fireball' should be the highest. Therefore, it is worthwhile to examine the behavior of flow parameters in a baryon rich environment, and compare them with those obtained at a high energy density and low baryon density condition prevalent in RHIC and LHC experiments. For the sake of completeness we have studied a few general features of multiparticle emission data, examined the flow parameters for different hadron species, and verified the NCQ scaling. The paper is organized as follows. In Sec.\,\ref{ampt} we summarize the AMPT model, in Sec.\,\ref{result} we sequentially describe the results obtained from this analysis followed by some discussions, and finally in Sec.\,\ref{summary} we conclude with a brief summary of our observations. 
\section{The AMPT Model}
\label{ampt}
As we intend to investigate the dependence of several observables on parton-parton scattering cross-section, it is necessary to choose an event generator that has an inbuilt provision for partonic degrees of freedom. AMPT is a hybrid transport model consisting of four major components namely, the initial conditions, the partonic interactions, the conversion from partonic to hadronic matter, and finally the hadronic interactions \cite {AMPT}. In AMPT the initial conditions are obtained through two body nucleon-nucleon $(NN)$ interactions, it uses a Glauber formalism to determine the positions of participating nucleons, and generates hard mini-jets (partons) and soft excited strings (hadrons) by using the heavy ion jet interaction generator (HIJING) \cite{HIJING}. The AMPT model can be used in two configurations, the default version and the string melting version. The basic difference between these two versions lies in modeling the excited strings. In the string melting mechanism beyond a certain critical energy density, excited strings (hadrons) and mini-jets (partons) cannot coexist. Therefore, it is necessary to melt or convert the strings into partons i.e., a meson is converted into a quark anti-quark pair, baryons into three quarks etc. The scattering among quarks and the original hard partons are then described by Zhang's Parton Cascade (ZPC) model \cite{Zhang98}, which includes two body elastic scattering with an in medium cross-section obtained from perturbative QCD (pQCD), where the effective gluon screening mass is used as a parameter. After the binary collisions cease to progress, the partons from mini-jets and partons from melted strings hadronize through a quark coalescence mechanism. However, in the AMPT default mode the energy of the excited strings are not used in the partonic stage. The scattering occurs only among the mini-jet partons based on the ZPC model and their hadronization is described by the Lund string fragmentation mechanism. After hadronization, either in the string melting version or in the default version, the hadron dynamics is modeled by a relativistic transport (ART) model \cite{Li95}, which includes both elastic and inelastic scatterings of baryonic, mesonic, and baryo-mesonic nature. Previous calculations have shown that flow parameters consistent with experiment, can be developed through AMPT, and the model can successfully describe different aspects of collective behavior of $AB$ interactions \cite{AMPT1, AMPT2, Partha10}. The string melting version of AMPT should be even more appropriate to model particle emission data where a transition from nuclear matter to deconfined QCD state is expected. We have used the AMPT model (string melting version) to generate $10^6$ minimum bias fixed target Au + Au interactions at $E_{\rm lab}$ = $30A$ GeV.
\begin{figure}[t]
\centering
\includegraphics[width=0.45\textwidth]{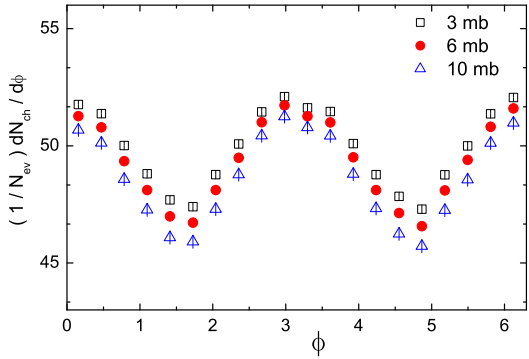}
\vspace{-.5cm}
\caption{(Color online) Azimuthal angle (measured in radian) distribution of charged hadrons produced in Au + Au collision at $E_{\rm lab}=30A$ GeV.}
\vspace{-0.2cm}
\label{azi}
\end{figure}
\section{Results}
\label{result}
\begin{figure}[t]
\centering
\includegraphics[width=0.45\textwidth]{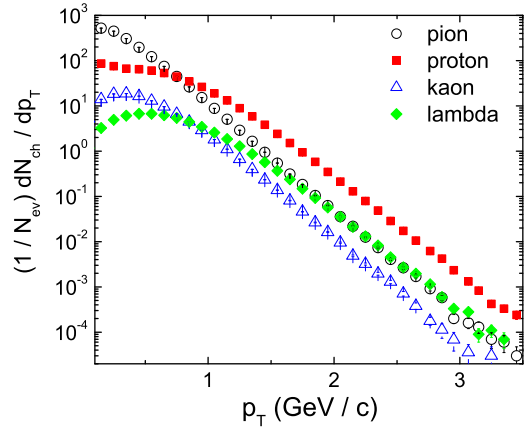}
\vspace{-.5cm}
\caption{(Color online) $p_{_T}$ distribution of identified charged hadrons produced in Au + Au collision at $E_{\rm lab}=30A$ GeV.}
\vspace{-0.2cm}
\label{hadron-pt}
\end{figure}

\begin{figure}[tbh]
\centering
\includegraphics[width=0.45\textwidth]{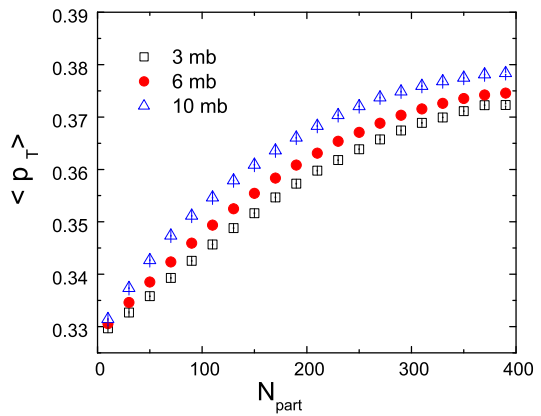}
\vspace{-.5cm}
\caption{(Color online) Average $p_{_T}$ of inclusive charged hadrons produced in Au + Au collision at $E_{\rm lab}=30A$ GeV plotted against centrality.}
\vspace{-0.2cm}
\label{npart-pT}
\end{figure}

In this section we describe the results obtained by analyzing our Au + Au minimum bias event sample at $E_{\rm lab}=30A$ GeV simulated by AMPT (String melting) model. Unless otherwise specified our results pertain to all charged hadrons produced in the Au + Au collisions. To begin with the azimuthal angle distributions are schematically plotted in Fig.\,\ref{azi} for three different two-body (partonic) scattering cross-sections $(\sigma)$. The $\sigma$ values are so chosen as to match the relevant and previously studied high-energy $AB$ interactions \cite{Han11, Partha10}. For $\sigma=3\;(10)$ mb. the number density of charged hadrons is found to be highest (lowest). Prominent anisotropies are observed in all three distributions that need to be investigated further. In Fig.\,\ref{hadron-pt} we have plotted the $p_{_T}$ distributions of some identified hadrons that come out of the AMPT generated Au + Au events at $E_{\rm lab}=30$A GeV. We observe the expected exponential decay in the distribution with increasing $p_{_T}$. The thermal region, i.e. the straight portion in the semi-log plot of $p_{_T}$ distribution for each hadron species, which in the present case roughly is $1.5 \leq p_{_T} \leq 2.5$ GeV/c, is fitted with an $dN_{\rm ch}/dp_{_T} \sim \exp(-\beta\,p_{_T})$ type of function. The values of fit parameter $\beta$ are $2.344\pm0.007$, $2.339\pm0.014$, $2.113\pm0.003$, and $2.063\pm0.007$, respectively for pions, kaons, protons, and lambda particles, indicating the flatness or steepness of the corresponding distribution. Hydrodynamics predicts that due to a collective radial motion, heavier particles gain more in $p_{_T}$ leading thereby to a flattening in the corresponding $p_{_T}$ spectrum. As a consequence, for heavier particles at low $p_{_T}$ reduced $v_2$ values are expected and the rise of $v_2$ with $p_{_T}$ should shift towards larger $p_{_T}$. In Fig.\,\ref{npart-pT} the average transverse momentum $\left<p_{_T}\right>$ is plotted against the centrality measure $N_{\rm part}$ of the collision. We see that after an initial linear rise, $\left<p_{_T}\right>$ value saturates with increasing centrality beyond $N_{\rm part}\approx 300$. In conformity with our expectation, $\left<p_{_T}\right>$ values lie within a narrow range between $0.33$ and $0.38$ GeV/c, with a mean lying somewhere around $0.35$ GeV/c. At all centrality we observe that a higher $\sigma$ consistently results in a higher $\left<p_{_T}\right>$, indicating thereby that the chance of binary interaction positively influences the extent to which transverse degrees of freedom are excited in the intermediate `fireball'. In \cite{Sarkar} we have attributed a higher saturation value of $\left<p_{_T}\right>$ to a higher isotropic radial flow of charged hadrons in the transverse plane. 
\subsection{Dependence on initial geometry}
\begin{figure}[t]
\centering
\includegraphics[width=0.45\textwidth]{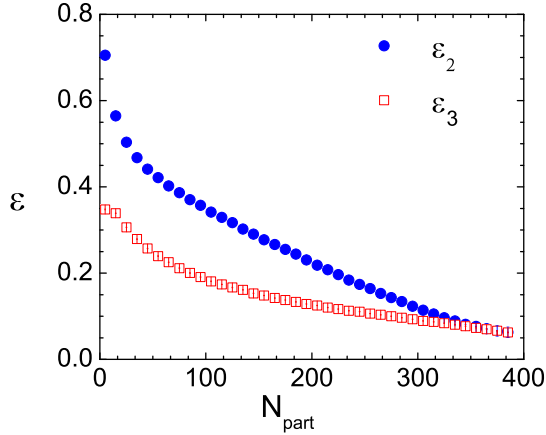}
\vspace{-.5cm}
\caption{(Color online) Centrality dependence of $\varepsilon_n$ of the overlapping region of Au + Au collision at $E_{\rm lab}=30A$ GeV.}
\vspace{-0.2cm}
\label{epsilon}
\end{figure}

\begin{figure}[t]
\centering
\hspace{-1.3cm}
\includegraphics[width=0.55\textwidth]{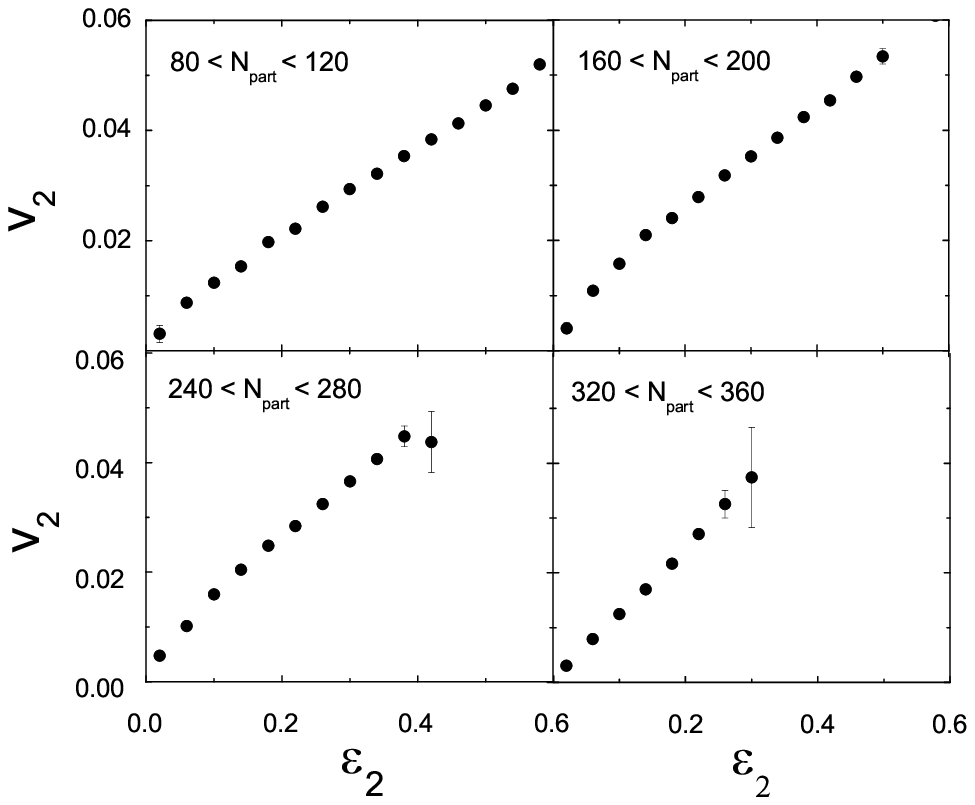}
\vspace{-.5cm}
\caption{$v_2$ as a function $\varepsilon_2$ in different $N_{\rm part}$ intervals for Au + Au collision at $E_{\rm lab}=30A$ GeV with $\sigma=3$ mb.}
\vspace{-0.2cm}
\label{v2-e}
\end{figure}

\begin{figure}[tbh]
\centering
\hspace{-1.3cm}
\includegraphics[width=0.5\textwidth]{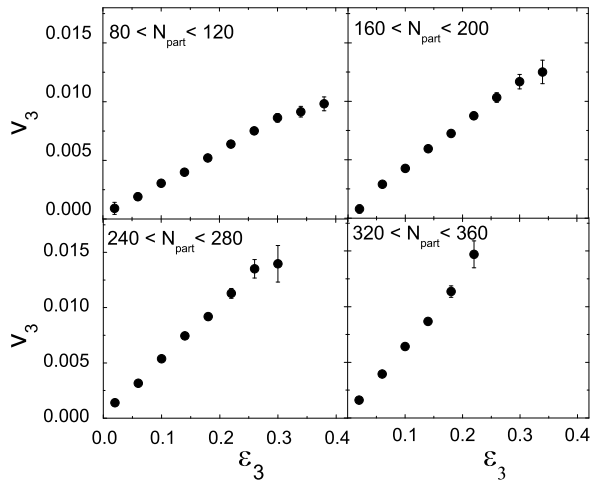}
\vspace{-.5cm}
\caption{$v_3$ as a function $\varepsilon_3$ in different $N_{\rm part}$ intervals for Au + Au collision at $E_{\rm lab}=30A$ GeV with $\sigma=3$ mb.}
\vspace{-0.2cm}
\label{v3-e}
\end{figure}

In Fig.\,\ref{epsilon} we present the centrality dependence of the initial geometric eccentricity $(\varepsilon_2)$ and triangularity $(\varepsilon_3)$ in the mid-rapidity region $(0\leq y \leq 4)$ obtained through Eq.\,(\ref{en}). As expected $\varepsilon_n$ decreases with $N_{\rm part}$. It should be noted that $\varepsilon_2$ is always greater than $\varepsilon_3$ except for the highest centrality region where they merge with each other. Taking the initial fluctuations into account [Eq.\,(\ref{vn})] we have calculated the elliptic $(v_2)$ and triangular $(v_3)$ flow parameters, and in Fig.\,\ref{v2-e} and Fig.\,\ref{v3-e} plotted their average values, respectively, as functions of eccentricity and triangularity in four different intervals of centrality. Extreme central and peripheral collisions are kept out of the purview of this part of the analysis. It is observed that both $v_2$ and $v_3$ increase with the corresponding geometric measure of anisotropy of the overlapping part of the colliding nuclei. However, with increasing $\varepsilon_2$ the rise in $v_2$ is steeper than that of $v_3$ with increasing $\varepsilon_3$. As expected, this is an indication that the efficiency with which initial spatial anisotropy gets converted into final state momentum space anisotropy, is more in elliptic flow than in the triangular flow. One should however, keep it in mind that the latter is not a consequence of any dynamics, but merely an outcome of initial fluctuations present in the distribution of participating nucleons in coordinate space. It is also interesting to note that for the four centrality intervals considered, the $v_n-\varepsilon_n~(n\,=\,2,\;3)$ dependence becomes steeper with increasing centrality, an observation which is almost similar to that of RHIC \cite{Alver10a}.\\
\subsection{Dependence on parton scattering cross-section}
\begin{figure}[t]
\centering
\vspace{-0.7cm}
\includegraphics[width=0.5\textwidth]{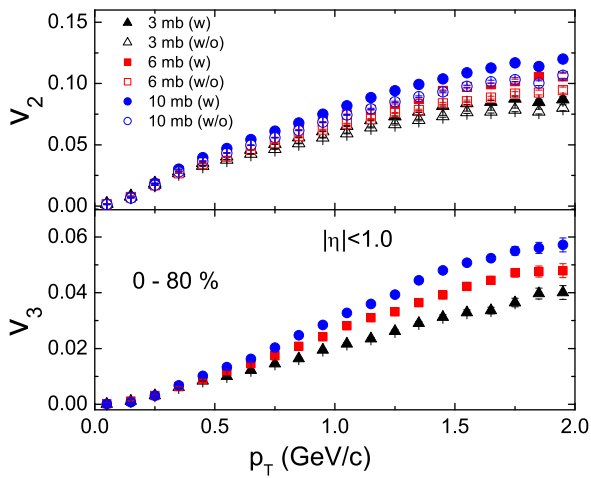}
\vspace{-.5cm}
\caption{(Color online) $v_2$ and $v_3$ as functions of $p_{_T}$ at mid-rapidity for Au + Au collision at $E_{\rm lab}=30A$ GeV.}
\vspace{-0.2cm}
\label{v2-v3-pt}
\end{figure}

\begin{figure}[t]
\centering
\includegraphics[width=0.5\textwidth]{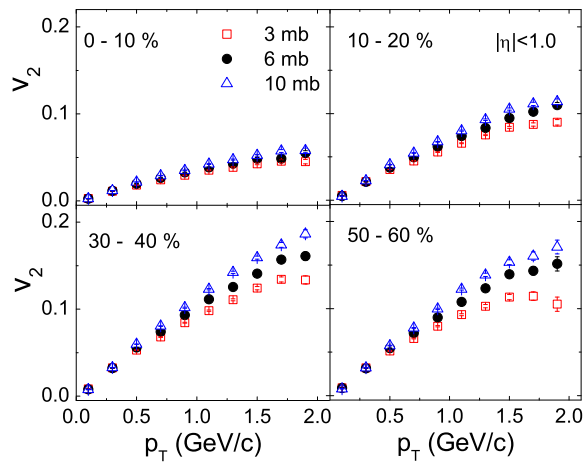}
\vspace{-.5cm}
\caption{(Color online) $p_{_T}$ dependence of $v_2$ for different centrality windows at mid-rapidity for Au + Au collision at $E_{\rm lab}=30A$ GeV.}
\vspace{-0.2cm}
\label{v2-cent}
\end{figure}

\begin{figure}[tbh]
\centering
\includegraphics[width=0.5\textwidth]{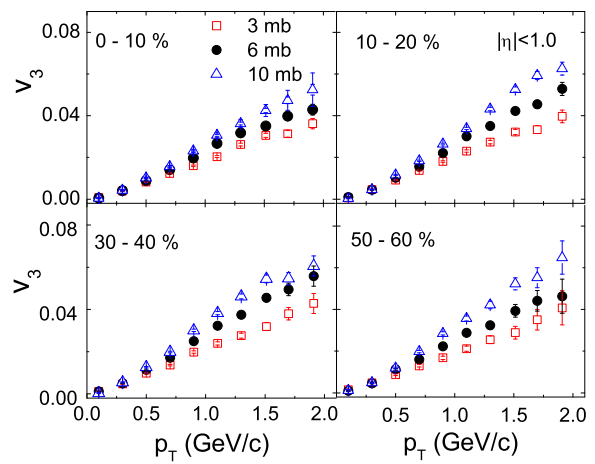}
\vspace{-.5cm}
\caption{(Color online) $p_{_T}$ dependence of $v_3$ for different centrality windows at mid-rapidity for Au + Au collision at $E_{\rm lab}=30A$ GeV.}
\vspace{-0.2cm}
\label{v3-cent}
\end{figure}

Fig.\,\ref{v2-v3-pt} shows the $p_{_T}$ dependence of differential elliptic flow and triangular flow parameters at mid-rapidity for different partonic scattering cross-sections. One should note that the $p_{_T}$ dependence of $v_2$ at a particular $\sigma$ has been presented both with (w) and without (w/o) considering the initial fluctuations in the position coordinates of the participating nucleons. For all $\sigma$ the $v_2$ values increase with increasing $p_{_T}$ and saturate at high $p_{_T}$. It should be noted that their is a small but definite positive impact of initial fluctuations on $v_2$ at all $\sigma$, which grows with increasing $p_{_T}$ as well as with increasing $\sigma$. It is also obvious that $v_3$ arises from the event-by-event fluctuations present in the initial collision geometry of the system, and the pattern (not the value) of its dependence on $p_{_T}$ is nearly the same as that of $v_2$. Once again a higher $\sigma$ results in a higher triangular flow. A consistently higher magnitude of $v_2$ over $v_3$ may be attributed to the fact that, while the former arises from the geometrical asymmetry of the overlapping region as well as from initial fluctuations, the latter results only from initial fluctuations. Obviously in comparison with the azimuthal asymmetry and the pressure gradient built thereof, the initial state fluctuation is a much weaker phenomenon. Further Fig.\,\ref{v2-v3-pt} helps us understand that the conversion efficiency from coordinate space anisotropy to momentum space anisotropy, grows with increasing $\sigma$ which in turn raises the magnitude of both $v_2$ and $v_3$. It is also to be understood from Fig.\,\ref{v2-v3-pt} that with increasing $p_{_T}$ the relative increase in $v_3$ values are more than those of $v_2$. Now, we intend to study the centrality dependence of $v_2$ and $v_3$ at central rapidity region. In Fig.\,\ref{v2-cent} we see that at low $p_{_T}$ for most central collisions $v_2$ is almost independent of $\sigma$.
However, the $\sigma$ dependence becomes prominent as we approach towards peripheral collisions. It is to be noted that $v_2$ values are maximum in mid-central region, where the dependence on partonic cross-section at high $p_{_T}$ is also maximum. For mid-central collisions we see a saturation  and for $\sigma=3$ mb. even a decreasing trend in $v_2$ at high $p_{_T}$. The observations supplement our result shown in Fig.\,\ref{v2-v3-pt}. In Fig.\,\ref{v3-cent} we have studied a similar centrality dependence of $v_3$. It is noticed that at each centrality bin considered, the variation of $v_3$ with $p_{_T}$ is more or less similar to that of $v_2$. In other words, the triangular flow is less sensitive to centrality. This is an expected result as the elliptic flow is caused by the pressure gradient created over the almond shape of the overlapping part of a collision, but triangular flow is caused from the fluctuations present in this shape. A smaller eccentricity of the overlapping part results in a higher pressure gradient and hence a larger value of elliptic flow. However, it is not necessary for the initial fluctuations to increase or decrease with centrality percentage.\\
\begin{figure}[tbh]
\centering
\includegraphics[width=0.5\textwidth]{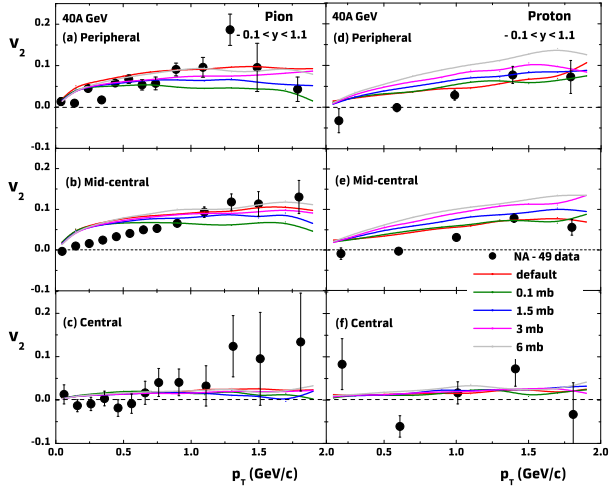}
\vspace{-.5cm}
\caption{(Color online) NA49 data on $p_{_T}$ dependence of $v_2$ obtained from Pb+Pb interactions at 40A GeV compared with AMPT simulation (both default and sring melting) for pions and protons at different partonic cross-sections and at different centralities. The experimental values are shown as points, while corresponding simulations are shown by continuous curves.}
\vspace{-0.2cm}
\label{NA49}
\end{figure}
\\
Before proceeding further it would perhaps be prudent to compare our AMPT simulation with existing experimental result(s) on elliptic flow at comparable energy. For this purpose we have chosen the NA49 experiment on Pb + Pb collisions at $E_{lab}=40A$ GeV \cite{NA49}. It has to be mentioned that in \cite{NA49} the standard $n$-th co-efficient of anisotropy has been evaluated by using a formula
\beqn
v_n = \frac{\left<\cos[n(\phi - \Phi_n)]\right>}{\left <\cos[n(\Phi_n - \Phi_{R})]\right>}
\label{NA49eq}
\eeqn
that is different from Eq.\,(\ref{vn}) used by us. In Eq.\,(\ref{NA49eq}) $\Phi_n$ represents the azimuthal angle of the event plane as explained in \cite{NA49}, and $\Phi_R$ represents that of the reaction plane i.e., the azimuthal angle of the impact parameter $b$. In this work we are interested only to investigate the effects arising out of fluctuating number of participating nucleons from one event to the other. While simulating the data we have not therefore, taken the changes in the orientation of the impact parameter into account. While determining $v_n$ we do not have to therefore, consider the event plane or its fluctuation. Moreover, in \cite{Poskanzer98} it has been shown that $\Phi_n\approx \Phi_R$. For our simulated data Eq.\,(\ref{NA49eq}) therefore, reduces simply to
\beqn
v_n = \left<\cos(n\phi)\right>
\label{NA49eqa}
\eeqn
In Fig.\,\ref{NA49} we have plotted $v_2$ against $p_{_T}$ for charged pions and protons separately as obtained from the NA49 experiment \cite{NA49}. Corresponding AMPT simulated values are also shown in the graph within the same $p_{_T}$ and same rapidity range as that of the experiment, as well as using the same centrality criteria as those used in \cite{NA49}. In spite of using a reasonably wide range of $\sigma$ values ($\sigma = 0.1,\, 1.5,\, 3.0,\, 6.0$ mb.), we observe that neither the default version of AMPT, nor the AMPT (string melting) version can match the entire set of experimental results for any single partonic cross-section. For soft hadrons ($p_{_T} < 1.0$ GeV/c) the experimental points behave in a fairly regular manner at least in the peripheral (more than $33.5\%$ centrality) and in mid-central ($12.5 - 33.5\%$ centrality) collisions. The AMPT however, exceeds the experiment in this region for all partonic cross-sections used in the SM version, and even in the default version. It is to be noted that each 40A GeV Au + Au simulated event sample used in this context has same statistics (i.e. $10^6$ Au + Au minimum bias events) as that used for the 30A GeV Au + Au simulated samples used in this paper. The disagreement at low $p_{_T}$ between experiment and simulation is more prominent in mid-central collisions for pions, and in peripheral as well as in mid-central collisions for protons. The percentage errors (statistical only) associated with the simulated $v_2$ values for pions in the $p_{_T} < 1.0$ GeV/c range are like, $< 4\%$ in peripheral and $< 3\%$ in mid-central collisions. Corresponding errors for protons are $<9\%$ in peripheral and $<5\%$ in mid-central collisions. On the other hand, in most central collisions the experimental values at high $p_{_T}$ are associated with large errors, and more than one simulation lines pass through them. In order to match the experiment with simulation, either the model perhaps requires a fine tuning, or to reduce errors there has to be experiments with higher statistics. The CBM experiment is expected generate much larger statistics than the NA49 experiment, and it would therefore, be interesting to see to what extent the flow results of AMPT simulation can come into agreement with the CBM experiment.
\begin{figure}[t]
\centering
\includegraphics[width=0.5\textwidth]{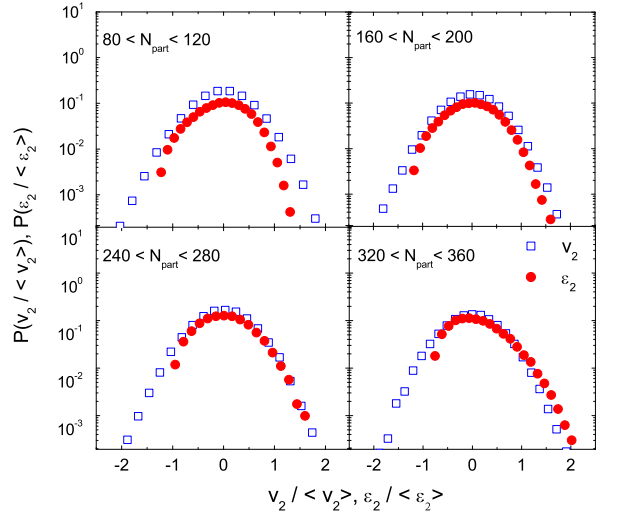}
\vspace{-.5cm}
\caption{(Color online) Distributions of $v_2/<v_2>$ and $\epsilon_2/<\epsilon_2>$ for charged hadrons in Au + Au collision at $E_{\rm lab}=30A$ GeV.}
\vspace{-0.2cm}
\label{v2-e2}
\end{figure}
\begin{figure}[tbh]
\centering
\includegraphics[width=0.5\textwidth]{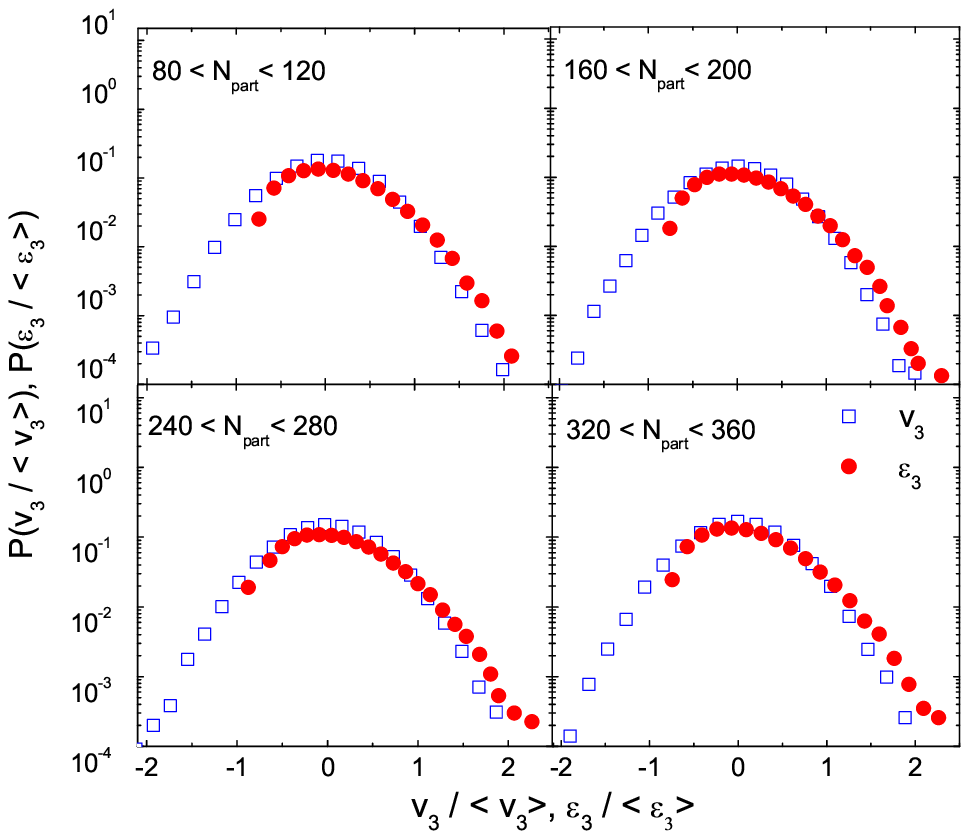}
\vspace{-.5cm}
\caption{(Color online) Distributions of $v_3/<v_3>$ and $\epsilon_3/<\epsilon_3>$ for charged hadrons in Au + Au collision at $E_{\rm lab}=30A$ GeV.}
\vspace{-0.2cm}
\label{v3-e3}
\end{figure}
\begin{figure}[t]
\centering
\includegraphics[width=0.5\textwidth]{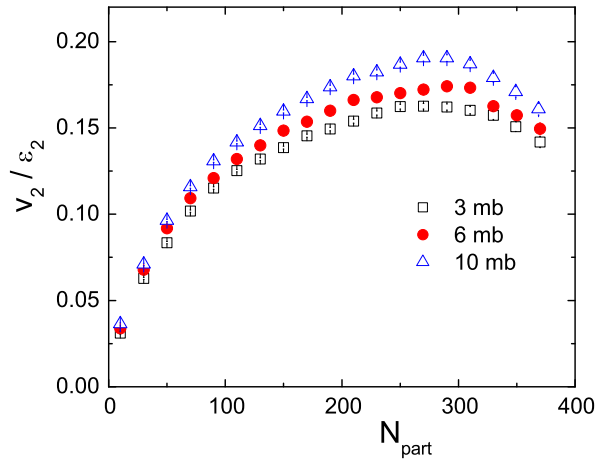}
\vspace{-.5cm}
\caption{(Color online) Elliptical flow scaled by eccentricity against centrality for Au + Au collision at $E_{\rm lab}=30A$ GeV.}
\vspace{-0.2cm}
\label{v2bye2}
\end{figure}
\begin{figure}[tbh]
\centering
\includegraphics[width=0.5\textwidth]{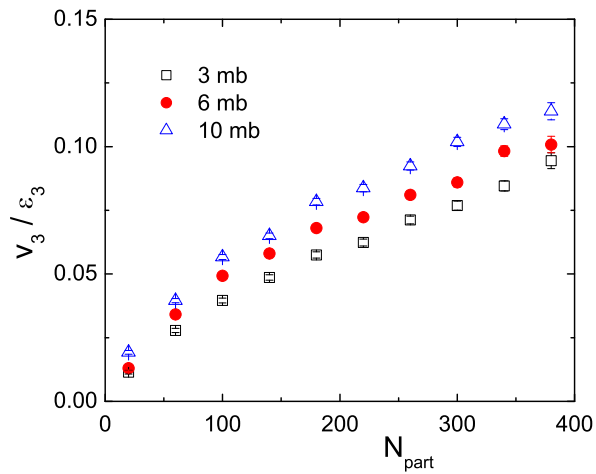}
\vspace{-.5cm}
\caption{(Color online) Triangular flow scaled by triangularity plotted against centrality for Au + Au collision at $E_{\rm lab}=30A$ GeV.}
\vspace{-0.2cm}
\label{v3bye3}
\end{figure}
\begin{figure}[tbh]
\centering
\includegraphics[width=0.5\textwidth]{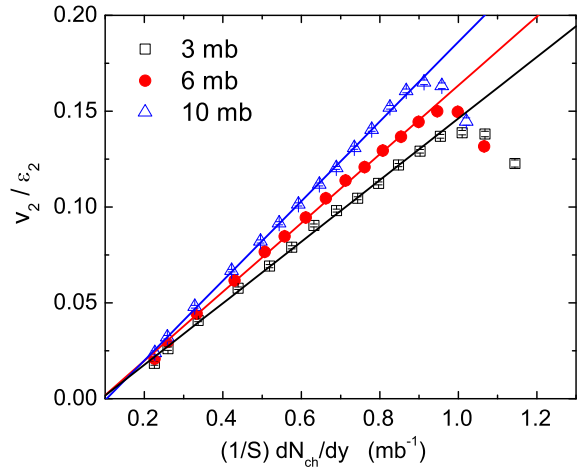}
\vspace{-.5cm}
\caption{(Color online) Elliptical flow scaled by eccentricity plotted against particle density in the transverse plane for Au + Au collision at $E_{\rm lab}=30A$ GeV. Solid lines represent best fits to the linear portion of the data.}
\vspace{-0.2cm}
\label{v2e-dndy}
\end{figure}
\\
\\ 
The probability distributions of asymmetry parameters obtained event wise, are now schematically represented in Fig.\,\ref{v2-e2} and in Fig.\,\ref{v3-e3} for the same four centrality intervals that are chosen before in this analysis, but only for one $\sigma$ ($3$ mb) \cite{Gale13}. The event to event fluctuations in participating nucleons are taken into consideration. To compare these distributions with the corresponding eccentricity distributions, both $v_n$ and $\epsilon_n$ are first scaled by their respective mean values, and then these scaled variables are converted to the respective standard normal variables. A strict proportionality like $v_n \propto \epsilon_n$ should result in a complete overlapping of the $P(v_n/<v_n>)$ and $P(\epsilon_n/<\epsilon_n>)$ distributions. However, both for $n=2$ and $3$ such overlapping can be seen only in limited regions. Significant differences between distributions of asymmetry and eccentricity parameters are seen in the most peripheral event sample $80\leq N_{\rm part} \leq 120$. We however see less mismatch between $v_3$ and $\epsilon_3$ distributions than that between $v_2$ and $\epsilon_2$. The $v_3$ distributions are consistently wider than the $v_2$ distributions. While both the flow parameters are almost symmetrically (normally) distributed, the eccentricity distributions are visibly skewed.\\
\\   
In Fig.\,\ref{v2bye2} and Fig.\,\ref{v3bye3} we schematically represent the centrality dependence of $v_n/\varepsilon_n$, a ratio known to be related to the freeze-out temperature \cite{Derek}. Though in Fig.\,\ref{v2-e} we found that in limited centrality intervals $v_2$ is proportional to $\varepsilon_2$, the $v_2/\varepsilon_2$ ratio shows a lot of variation with varying centrality. We see that in the low centrality region the $v_2/\varepsilon_2$ ratio increases almost linearly with increasing centrality, gets nonlinear in the mid-central region, reaches a maximum at $N_{\rm part}\approx 250$, and finally drops down from its maximum point within a very small interval of very high centrality, where the spatial asymmetry of the almond shaped overlapping region of the colliding Au nuclei is vanishingly small. It has been argued that in the low density limit of the intermediate `fireball' created in $AB$ collisions, the elliptic flow should be proportional to the elliptic anisotropy and the initial particle density \cite{Voloshin}, which certainly is not the case for our analysis. The eccentricity scaled elliptic flow is highest at the highest $\sigma$ considered in this analysis. On the contrary, the triangular flow parameter when scaled by the corresponding triangularity, increases monotonically (almost linear) with increasing centrality. Once again a higher $\sigma$ consistently results in a higher scaled triangular flow. It appears that experimentally obtained $v_n/\varepsilon_n$ ratio can perhaps be modeled by suitably adjusting $\sigma$ as a parameter. To further verify the behavior of scaled elliptical flow under the low density limit, in Fig.\,\ref{v2e-dndy} we have plotted the $v_2/\varepsilon_2$ ratio against the particle density in transverse plane. Once again it is found that except for a few very high centrality intervals, the proportionality 
\beqn
\frac{v_2}{\varepsilon_2}\, \propto \, \frac{1}{S}\,\frac{dN_{\rm ch}}{dy}
\label{scaling}
\eeqn
holds good. The proportionality constant may depend on the hydro limit of $v_2/\varepsilon$, binary scattering cross-section, and velocity of elastic wave in the medium concerned \cite{Drescher}. Here $S$ is the transverse area of the overlapping zone of the colliding nuclei, and $dN_{\rm ch}/dy$ is the rapidity density (a measure of rescattering within the `fireball') of charged hadrons. A higher $\sigma$ corresponds to a higher slope of the linear relationship as prescribed in Eq.\,(\ref{scaling}). At a few extreme high centralities the observed sudden deviation from the linear rising trend of the rest, may be attributed to a different physics associated with the corresponding `fireball' medium, which is potentially an interesting issue that needs further scrutiny.
\subsection{Relative strength of $v_2$ and $v_3$}
\begin{figure}[t]
\centering
\includegraphics[width=0.45\textwidth]{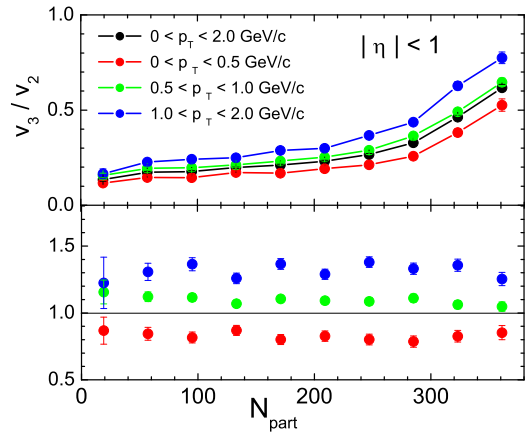}
\vspace{-.5cm}
\caption{(Color online) $\left<v_3\right>/\left<v_2\right>$ ratio plotted as a function of $N_{\rm part}$ in different $p_{_T}$ intervals for Au + Au collision at $E_{\rm lab}=30A$ GeV (upper panel). The same plot, but now each $\left<v_3\right>/\left<v_2\right>$ value in different $p_{_T}$ intervals is scaled by the corresponding ratio for the entire ($0\leq p_{_T}\leq 2.0$ GeV/c) interval (lower panel).}
\vspace{-0.2cm}
\label{ratio}
\end{figure}

\begin{figure}[t]
\centering
\includegraphics[width=0.5\textwidth]{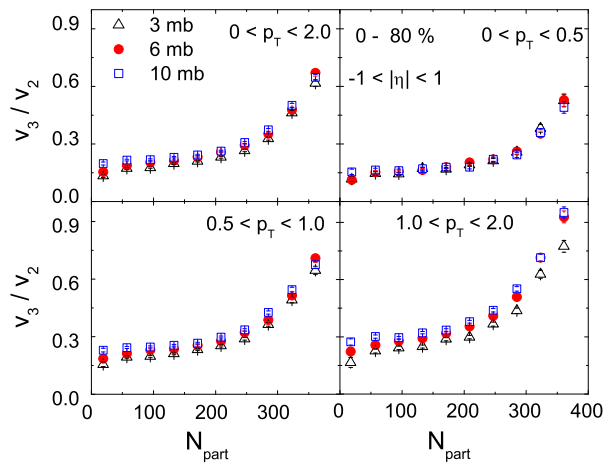}
\vspace{-.5cm}
\caption{(Color online) Dependence of ${v_3}/{v_2}$ on $N_{\rm part}$ in different $p_{_T}$ bins for Au + Au collision at $E_{\rm lab}=30A$ GeV.}
\vspace{-0.2cm}
\label{ratio-crs}
\end{figure}

We also compute the relative magnitude of the triangular flow with respect to the elliptic flow as a function of $N_{\rm part}$ in different $p_{_T}$ intervals and within $\Delta\eta=\pm1.0$ about the central $\eta$ value of the distribution. In Fig.\,\ref{ratio} the relative strength of $v_3$ is observed to increase with centrality, the rate of increase gets higher at the highest centrality region. The results shown in Fig.\,\ref{ratio} pertain to $\sigma = 3$ mb. Similar analysis however, is also performed at $\sigma$ = $6$ and $10$ mb. The gross features of ${v_3}/{v_2}$ ratio as a function of $N_{\rm part}$ are found to be more or less similar at all $\sigma$ values considered. The $v_3/v_2$ ratio is consistently higher at higher $p_{_T}$. However, when we divide the $v_3/v_2$ ratio corresponding to a particular $p_{_T}$ interval with that of the entire $p_{_T}$ interval, which in our case is $0\leq p_{_T}\leq 2.0\;\mbox{GeV/c}$, we find that the $v_3/v_2$ ratio so normalized becomes almost independent of centrality within statistical uncertainties. This result is schematically presented in the lower panel of Fig.\,\ref{ratio}, and our observation in this regard is similar to that of an AMPT simulation of RHIC experiment \cite{Alver10a}. When we plot the same ratio against $N_{\rm part}$ in different $p_{_T}$ intervals but for different $\sigma$ [Fig.\,\ref{ratio-crs}], we see that the relative magnitude $v_3/v_2$ initially remains almost constant, and then increases nonlinearly with increasing centrality following almost a power law. At low $p_{_T}$ $(\le 0.5\;\mbox{GeV/c})$ as well as for the entire $p_{_T}$ range, however, it is more or less independent of $\sigma$. As shown in the lower panels of Fig.\,\ref{ratio-crs} the $\sigma$ dependence of $v_3/v_2$ increases marginally at high $p_{_T}$ range.
\subsection{Species dependence and NCQ scaling}
\begin{figure}[t]
\centering
\vspace{-0.7cm}
\includegraphics[width=0.5\textwidth]{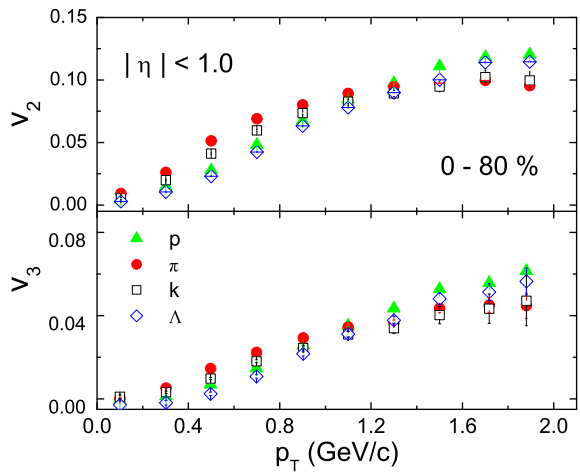}
\vspace{-0.5cm}
\caption{(Color online) Species dependence of $v_2$ (upper panel) and $v_3$ (lower panel) as a function of $p_{_T}$ for Au + Au collision at $E_{\rm lab}=30A$ GeV. }
\vspace{-0.2cm}
\label{mass-ordering}
\end{figure}

An important aspect of azimuthal anisotropy is the mass ordering of flow parameters of identified hadron species produced in high-energy $AB$ interactions. In Fig.\,\ref{mass-ordering} we present the $v_2$ and $v_3$ values plotted against $p_{_T}$ at mid-rapidity for $0 - 80 \%$ centrality range for different species of produced hadrons tuning our simulated data to $\sigma = 3$ mb. Below $p_{_T} \approx 1.1$ GeV/c both the elliptic and the triangular flow parameters show an obvious mass ordering, i.e. higher $v_n$ for lower hadron mass, which is consistent with the hydrodynamic prediction. It is interesting to note that beyond $p_{_T}\approx 1.2$ GeV/c this mass ordering is no longer preserved. It actually gets inverted between mesons and baryons, and at $p_{_T} > 1.4$ GeV/c the mass ordering trends for baryons and mesons split into two separate bands. This feature can be ascribed to the fact that provided an extended QCD state is formed, both $v_2$ and $v_3$ are expected to depend on the constituent partonic degrees of freedom of respective baryon and meson species. As mentioned above, a redistribution of the momentum anisotropy will then built up due to a mass-dependent flattening of the $p_{_T}$ spectra caused by a radial flow generated during the hadronization process, thus resulting in the mass splitting. Similar mass ordering has been reported in RHIC \cite{Adler01, Abelev07,Adler02} and LHC experiments \cite{Abelev15}, and also in AMPT simulation of $AB$ collisions at RHIC and FAIR energies \cite{Han11, Partha10}. It is perhaps due to the quark coalescence mechanism there is a tendency, that the differential flow parameters pertaining to a particular hadron species (meson or baryon) group together. The recombination of constituent quarks neighboring each other in phase space, is also expected to lead to a uniform behavior in the way the flow parameters should depend on the transverse degrees of freedom. In particular, when appropriately scaled by the number of constituent quarks, hadrons belonging to different species are supposed to depend identically on $p_{_T}$. The phenomenon known as NCQ scaling \cite{Molnar03}, has been verified in RHIC experiments \cite{NCQ-RHIC}, is cosnidered to be an important evidence of partonic degrees of freedom present in the `fireball', and is an integrated consequnce of both partonic and hadronic interactions \cite{Zheng16}.
\begin{figure}[t]
\centering
\includegraphics[width=0.5\textwidth]{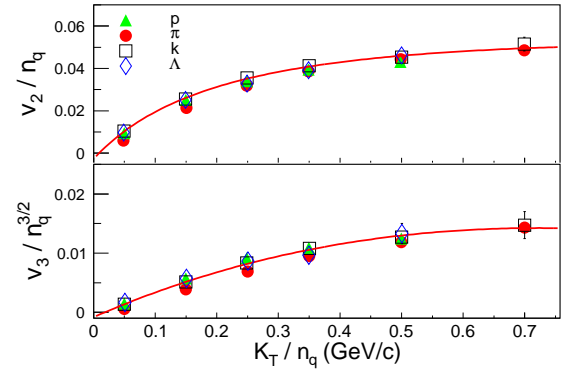}
\vspace{-.5cm}
\caption{(Color online) $v_2$ (upper panel) and $v_3$ (lower panel) scaled by the constituent quark numbers of hadrons as a function of $K_{_T}/n_q$ for Au + Au collision at $E_{\rm lab}=30A$ GeV.}
\vspace{-0.2cm}
\label{ncq-scaling}
\end{figure}
In Fig.\,\ref{ncq-scaling} we have shown the dependence of $v_2$ and $v_3$ on the transverse kinetic energy $K_{_T}=\sqrt{p_{_T}^2+m_0^2}-m_0$. Following the proposal made in \cite{Molnar03}, the $K_{_T}$ and $v_2$ values are scaled by $n_q$, while $v_3$ is scaled by $n_q^{3/2}$. In general $v_n$ has to be scaled by $n_q^{n/2}$, which specifies how partonic interactions differently influence the flow parameters pertaining to different harmonics. Within statistical uncertainties our result agrees reasonably well with NCQ. One may speculate that the collective behavior has developed quite early in the partonic stage of the fireball and it also corroborates a quark coalescence picture of hadronization. A phenomenologically motivated fit function of the form 
\beqn
\frac{v_n}{n_q^{n/2}} = \frac{a + b\,x + c\,x^2}{d\,-\,x} -\frac{a}{2}
\eeqn
where $x$ = $K_{_T}/n_q$, describes the scaling satisfactorily \cite{Sorensen}. However, at this stage we do not intend to assign any physical significance to the fit. 
\section{Summary}
\label{summary}
In the framework of the AMPT (string melting) model, in this paper we have presented some results on elliptical and triangular flow of charged hadrons produced in Au + Au interaction at $E_{\rm lab}=30A$ GeV. Dependence of flow parameters on initial conditions and binary (partonic) scattering cross-section are investigated. The major observations of this analysis are summarized below.
AMPT (string melting) version is capable of generating momentum space anisotropy even at FAIR energies. The dependence of both elliptical and triangular flow parameters on the corresponding geometrical asymmetry, transverse momentum of charged hadrons and centrality of collision, are of expected nature. Event to event initial fluctuations not only result in a non-zero triangular flow, but it also has a small but definite positive impact on the elliptic flow. The dependence of flow parameters on partonic scattering cross-section are almost always qualitatively similar but quantitatively different by a small amount. However, the NA49 experimental results, particularly at low $p_{_t}$ and in mid-central to peripheral collisions, could not be satisfactorily reproduced by the AMPT simulation either in its default or in the string melting version even though several different partonic cross-sections are used. While putting the CBM results to similar comparative test, one will have to be therefore, careful about the experimental conditions, the statistics, and the technique(s) of data analysis. Except for some low and very high values of anisotropy, to a good approximation the respective initial eccentricities are capable of representing the distributions of asymmetry parameters considered in this analysis. Both the mass ordering of flow parameters of charged hadrons belonging to different species, and the scaling with respect to their constituent quark numbers are observed. These simulated results will help us understand several issues related to the collective behavior of hadronic and/or partonic matter in a baryon rich and moderate temperature environment until real experiments are held.

\end{document}